\input epsf
%
%
%
\def\unredoffs{} 

%
%
%
%
\newbox\leftpage \newdimen\fullhsize \newdimen\hstitle \newdimen\hsbody
\tolerance=1000\hfuzz=2pt
\catcode`\@=11 
%
\magnification=1200\unredoffs\baselineskip=16pt plus 2pt minus 1pt
\hsbody=\hsize \hstitle=\hsize 
%
%
%
\newcount\yearltd\yearltd=\year\advance\yearltd by -1900

%
%

\def\draftmode{\message{ DRAFTMODE }\def\draftdate{{\rm preliminary draft:
\number\month/\number\day/\number\yearltd\ \ \hourmin}}%
\headline={\hfil\draftdate}\writelabels\baselineskip=20pt plus 2pt minus 2pt
 {\count255=\time\divide\count255 by 60 \xdef\hourmin{\number\count255}
  \multiply\count255 by-60\advance\count255 by\time
  \xdef\hourmin{\hourmin:\ifnum\count255<10 0\fi\the\count255}}}
\def\nolabels{\def\wrlabeL##1{}\def\eqlabeL##1{}\def\reflabeL##1{}}
\def\writelabels{\def\wrlabeL##1{\leavevmode\vadjust{\rlap{\smash%
{\line{{\escapechar=` \hfill\rlap{\sevenrm\hskip.03in\string##1}}}}}}}%
\def\eqlabeL##1{{\escapechar-1\rlap{\sevenrm\hskip.05in\string##1}}}%
\def\reflabeL##1{\noexpand\llap{\noexpand\sevenrm\string\string\string##1}}}
\nolabels
%
\global\newcount\secno \global\secno=0
\global\newcount\meqno \global\meqno=1
\def\newsec#1{\global\advance\secno by1\message{(\the\secno. #1)}
\global\subsecno=0\eqnres@t\noindent{\bf\the\secno. #1}
\writetoca{{\secsym} {#1}}\par\nobreak\medskip\nobreak}
\def\eqnres@t{\xdef\secsym{\the\secno.}\global\meqno=1\bigbreak\bigskip}
\def\sequentialequations{\def\eqnres@t{\bigbreak}}\xdef\secsym{}
\global\newcount\subsecno \global\subsecno=0
\def\subsec#1{\global\advance\subsecno by1\message{(\secsym\the\subsecno. #1)}
\ifnum\lastpenalty>9000\else\bigbreak\fi
\noindent{\it\secsym\the\subsecno. #1}\writetoca{\string\quad
{\secsym\the\subsecno.} {#1}}\par\nobreak\medskip\nobreak}
\def\appendix#1#2{\global\meqno=1\global\subsecno=0\xdef\secsym{\hbox{#1.}}
\bigbreak\bigskip\noindent{\bf Appendix #1. #2}\message{(#1. #2)}
\writetoca{Appendix {#1.} {#2}}\par\nobreak\medskip\nobreak}
%
%
\def\eqnn#1{\xdef #1{(\secsym\the\meqno)}\writedef{#1\leftbracket#1}%
\global\advance\meqno by1\wrlabeL#1}
\def\eqna#1{\xdef #1##1{\hbox{$(\secsym\the\meqno##1)$}}
\writedef{#1\numbersign1\leftbracket#1{\numbersign1}}%
\global\advance\meqno by1\wrlabeL{#1$\{\}$}}
\def\eqn#1#2{\xdef #1{(\secsym\the\meqno)}\writedef{#1\leftbracket#1}%
\global\advance\meqno by1$$#2\eqno#1\eqlabeL#1$$}
%
\newskip\footskip\footskip14pt plus 1pt minus 1pt 
\def\footnotefont{\ninepoint}\def\f@t#1{\footnotefont #1\@foot}
\def\f@@t{\baselineskip\footskip\bgroup\footnotefont\aftergroup\@foot\let\next}
\setbox\strutbox=\hbox{\vrule height9.5pt depth4.5pt width0pt}
\global\newcount\ftno \global\ftno=0
\def\foot{\global\advance\ftno by1\footnote{$^{\the\ftno}$}}
%
\newwrite\ftfile
\def\footend{\def\foot{\global\advance\ftno by1\chardef\wfile=\ftfile
$^{\the\ftno}$\ifnum\ftno=1\immediate\openout\ftfile=foots.tmp\fi%
\immediate\write\ftfile{\noexpand\smallskip%
\noexpand\item{f\the\ftno:\ }\pctsign}\findarg}%
\def\footatend{\vfill\eject\immediate\closeout\ftfile{\parindent=20pt
\centerline{\bf Footnotes}\nobreak\bigskip\input foots.tmp }}}
\def\footatend{}
%
%
\global\newcount\refno \global\refno=1
\newwrite\rfile
\def\ref{[\the\refno]\nref}
\def\nref#1{\xdef#1{[\the\refno]}\writedef{#1\leftbracket#1}%
\ifnum\refno=1\immediate\openout\rfile=refs.tmp\fi
\global\advance\refno by1\chardef\wfile=\rfile\immediate
\write\rfile{\noexpand\item{#1\ }\reflabeL{#1\hskip.31in}\pctsign}\findarg}
\def\findarg#1#{\begingroup\obeylines\newlinechar=`\^^M\pass@rg}
{\obeylines\gdef\pass@rg#1{\writ@line\relax #1^^M\hbox{}^^M}%
\gdef\writ@line#1^^M{\expandafter\toks0\expandafter{\striprel@x #1}%
\edef\next{\the\toks0}\ifx\next\em@rk\let\next=\endgroup\else\ifx\next\empty%
\else\immediate\write\wfile{\the\toks0}\fi\let\next=\writ@line\fi\next\relax}}
\def\striprel@x#1{} \def\em@rk{\hbox{}}
\def\lref{\begingroup\obeylines\lr@f}
\def\lr@f#1#2{\gdef#1{\ref#1{#2}}\endgroup\unskip}

\def\addref#1{\immediate\write\rfile{\noexpand\item{}#1}} 
\def\startrefs#1{\immediate\openout\rfile=refs.tmp\refno=#1}
\def\xref{\expandafter\xr@f}\def\xr@f[#1]{#1}
\def\refs#1{\count255=1[\r@fs #1{\hbox{}}]}
\def\r@fs#1{\ifx\und@fined#1\message{reflabel \string#1 is undefined.}%
\nref#1{need to supply reference \string#1.}\fi%
\vphantom{\hphantom{#1}}\edef\next{#1}\ifx\next\em@rk\def\next{}%
\else\ifx\next#1\ifodd\count255\relax\xref#1\count255=0\fi%
\else#1\count255=1\fi\let\next=\r@fs\fi\next}
%

%
\newwrite\ffile\global\newcount\figno \global\figno=1
\def\fig{fig.~\the\figno\nfig}
\def\nfig#1{\xdef#1{fig.~\the\figno}%
\writedef{#1\leftbracket fig.\noexpand~\the\figno}%
\ifnum\figno=1\immediate\openout\ffile=figs.tmp\fi\chardef\wfile=\ffile%
\immediate\write\ffile{\noexpand\medskip\noexpand\item{Fig.\ \the\figno. }
\reflabeL{#1\hskip.55in}\pctsign}\global\advance\figno by1\findarg}
\def\xfig{\expandafter\xf@g}\def\xf@g fig.\penalty\@M\ {}
\def\figs#1{figs.~\f@gs #1{\hbox{}}}
\def\f@gs#1{\edef\next{#1}\ifx\next\em@rk\def\next{}\else
\ifx\next#1\xfig #1\else#1\fi\let\next=\f@gs\fi\next}
\newwrite\lfile
{\escapechar-1\xdef\pctsign{\string\%}\xdef\leftbracket{\string\{}
\xdef\rightbracket{\string\}}\xdef\numbersign{\string\#}}

\def\writestop{\def\writestoppt{\immediate\write\lfile{\string\pageno%
\the\pageno\string\startrefs\leftbracket\the\refno\rightbracket%
\string\def\string\secsym\leftbracket\secsym\rightbracket%
\string\secno\the\secno\string\meqno\the\meqno}\immediate\closeout\lfile}}
\def\writestoppt{}\def\writedef#1{}
\def\seclab#1{\xdef #1{\the\secno}\writedef{#1\leftbracket#1}\wrlabeL{#1=#1}}
\def\subseclab#1{\xdef #1{\secsym\the\subsecno}%
\writedef{#1\leftbracket#1}\wrlabeL{#1=#1}}
\newwrite\tfile \def\writetoca#1{}
\def\leaderfill{\leaders\hbox to 1em{\hss.\hss}\hfill}
\def\writetoc{\immediate\openout\tfile=toc.tmp
   \def\writetoca##1{{\edef\next{\write\tfile{\noindent ##1
   \string\leaderfill {\noexpand\number\pageno} \par}}\next}}}

%
\catcode`\@=12 
%
\edef\tfontsize{\ifx\answ\bigans scaled\magstep3\else scaled\magstep4\fi}
\font\titlerm=cmr10 \tfontsize \font\titlerms=cmr7 \tfontsize
\font\titlermss=cmr5 \tfontsize \font\titlei=cmmi10 \tfontsize
\font\titleis=cmmi7 \tfontsize \font\titleiss=cmmi5 \tfontsize
\font\titlesy=cmsy10 \tfontsize \font\titlesys=cmsy7 \tfontsize
\font\titlesyss=cmsy5 \tfontsize \font\titleit=cmti10 \tfontsize
\skewchar\titlei='177 \skewchar\titleis='177 \skewchar\titleiss='177
\skewchar\titlesy='60 \skewchar\titlesys='60 \skewchar\titlesyss='60
\def\titlefont{\def\rm{\fam0\titlerm}
\textfont0=\titlerm \scriptfont0=\titlerms \scriptscriptfont0=\titlermss
\textfont1=\titlei \scriptfont1=\titleis \scriptscriptfont1=\titleiss
\textfont2=\titlesy \scriptfont2=\titlesys \scriptscriptfont2=\titlesyss
\textfont\itfam=\titleit \def\it{\fam\itfam\titleit}\rm}
 \ifx\answ\bigans\else scaled\magstep1\fi
\ifx\answ\bigans\else

 \font\absi=cmmi10 scaled\magstep1
\font\absis=cmmi7 scaled\magstep1 \font\absiss=cmmi5 scaled\magstep1
\font\abssy=cmsy10 scaled\magstep1 \font\abssys=cmsy7 scaled\magstep1
\font\abssyss=cmsy5 scaled\magstep1 
\skewchar\absi='177 \skewchar\absis='177 \skewchar\absiss='177
\skewchar\abssy='60 \skewchar\abssys='60 \skewchar\abssyss='60
\fi
\font\ninerm=cmr9 \font\sixrm=cmr6 \font\ninei=cmmi9 \font\sixi=cmmi6
\font\ninesy=cmsy9 \font\sixsy=cmsy6 \font\ninebf=cmbx9
\font\nineit=cmti9 \font\ninesl=cmsl9 \skewchar\ninei='177
\skewchar\sixi='177 \skewchar\ninesy='60 \skewchar\sixsy='60
\def\ninepoint{\def\rm{\fam0\ninerm}
\textfont0=\ninerm \scriptfont0=\sixrm \scriptscriptfont0=\fiverm
\textfont1=\ninei \scriptfont1=\sixi \scriptscriptfont1=\fivei
\textfont2=\ninesy \scriptfont2=\sixsy \scriptscriptfont2=\fivesy
\textfont\itfam=\ninei \def\it{\fam\itfam\nineit}\def\sl{\fam\slfam\ninesl}%
\textfont\bffam=\ninebf \def\bf{\fam\bffam\ninebf}\rm}
%
%
\def\noblackbox{\overfullrule=0pt}
\hyphenation{anom-aly anom-alies coun-ter-term coun-ter-terms}
\def\inv{^{\raise.15ex\hbox{${\scriptscriptstyle -}$}\kern-.05em 1}}

\def\Dsl{\,\raise.15ex\hbox{/}\mkern-13.5mu D} 
\def\dsl{\raise.15ex\hbox{/}\kern-.57em\partial}

\def\lspace{\ifx\answ\bigans{}\else\qquad\fi}
\def\lbspace{\ifx\answ\bigans{}\else\hskip-.2in\fi} 
\def\boxeqn#1{\vcenter{\vbox{\hrule\hbox{\vrule\kern3pt\vbox{\kern3pt
        \hbox{${\displaystyle #1}$}\kern3pt}\kern3pt\vrule}\hrule}}}
\def\mbox#1#2{\vcenter{\hrule \hbox{\vrule height#2in
                \kern#1in \vrule} \hrule}}  
%

\def\darr#1{\raise1.5ex\hbox{$\leftrightarrow$}\mkern-16.5mu #1}

\def\half{{\textstyle{1\over2}}} 
\def\roughly#1{\raise.3ex\hbox{$#1$\kern-.75em\lower1ex\hbox{$\sim$}}}
\hyphenation{Mar-ti-nel-li}

\def\1{\;1\!\!\!\! 1\;}

\def\frac#1#2{{{#1}\over {#2}}}
\def\half{\hbox{${1\over 2}$}}
\def\quarter{\hbox{${1\over 4}$}}
\def\smallfrac#1#2{\hbox{${{#1}\over {#2}}$}}

\catcode`@=11 
\def\slash#1{\mathord{\mathpalette\c@ncel#1}}
 \def\c@ncel#1#2{\ooalign{$\hfil#1\mkern1mu/\hfil$\crcr$#1#2$}}
\def\lsim{\mathrel{\mathpalette\@versim<}}
\def\gsim{\mathrel{\mathpalette\@versim>}}
 \def\@versim#1#2{\lower0.2ex\vbox{\baselineskip\z@skip\lineskip\z@skip
       \lineskiplimit\z@\ialign{$\m@th#1\hfil##$\crcr#2\crcr\sim\crcr}}}
\catcode`@=12 

\def\as{\alpha_s}

\noblackbox
\pageno=0\nopagenumbers\tolerance=10000\hfuzz=5pt
\baselineskip 12pt
\line{\hfill {\tt hep-ph/0310016}}
\line{\hfill CERN-TH/2003-238}
\line{\hfill Edinburgh 2003/18}
\line{\hfill IFUM-774/FT}
\vskip 12pt
\centerline{{\titlefont An Improved Splitting Function for Small $x$
Evolution}}
\vskip 36pt\centerline{Guido~Altarelli,$^{(a)}$
Richard D.~Ball$^{(b)}$ and Stefano Forte$^{(c)}$}
\vskip 12pt
\centerline{\it ${}^{(a)}$Theory Division, CERN}
\centerline{\it CH-1211 Gen\`eve 23, Switzerland}
\vskip 6pt
\centerline{\it ${}^{(b)}$School of Physics, University of Edinburgh}
\centerline{\it  Edinburgh EH9 3JZ, Scotland}
\vskip 6pt
\centerline {\it ${}^{(c)}$Dipartimento di  Fisica, Universit\`a di
Milano and}
\centerline{\it INFN, Sezione di Milano, Via Celoria 16, I-20133
Milan, Italy\foot{Permanent address}}
\centerline{\it and}
\centerline {\it Centre de Physique Th\'eorique, Ecole Polytechnique}
\centerline{\it F-91128 Palaiseau, France}
\vskip 50pt
\centerline{\bf Abstract}
We summarize our recent result for a splitting function for small $x$
evolution which includes resummed small $x$ logarithms deduced from
the leading order BFKL equation with the inclusion of running coupling
effects. We compare this improved splitting function 
with alternative approaches.
\vskip1cm
\centerline{\it Presented by G.A. at DIS 2003}
\centerline{\it St. Petersburg, April 2003}
\centerline{\it To be published in the proceedings}

{\narrower\baselineskip 10pt
\medskip\noindent 

}
\vfill
\line{CERN-TH/2003-238\hfill }
\line{September 2003\hfill}
\eject \footline={\hss\tenrm\folio\hss}

\lref\newabf{G.~Altarelli, R.~D.~Ball and S.~Forte,
{\tt hep-ph/0306156} and ref. therein.}

\lref\newciaf{M.~Ciafaloni, D.~Colferai, G.~P.~Salam and A.~M.~Stasto,
{\tt hep-ph/0307188} and ref. therein;
D.~Colferai, {\it these Proceedings}}

\noindent In recent years the theory of scaling violations for deep inelastic
structure functions at small $x$ has
attracted considerable interest, prompted by the
experimental information coming from
HERA. New effects beyond the low--order perturbative
approximation to
anomalous dimensions  or splitting functions should become important at
small-$x$.  However, no major deviation of
the data from a standard
next--to--leading order perturbative treatment of  scaling
violations has been found.
By now the origin of this situation has been
mostly
understood~\refs{\newabf,\newciaf}.

The BFKL kernel $\chi(\alpha_s,M)$ has been computed to
next-to-leading accuracy (NLO):
\eqn\chidef{
\chi(M,\alpha_s)=\alpha_s \chi_0(M)~+~\alpha_s^2 \chi_1(M)~+~\dots . } The
problem is how to use the information contained
in
$\chi_0$ and $\chi_1$ in order to improve the splitting function derived from
the perturbative leading singlet anomalous
dimension function $\gamma(\alpha_s,N)$ which is known up to
NLO in $\alpha_s$:
\eqn\gammadef{
\gamma(N,\alpha_s)=\alpha_s \gamma_0(N)~+~\alpha_s^2
\gamma_1(N)~+~\dots ,} in such a way that the improved splitting function
remains a good approximation down to small values
of $x$. This can be accomplished~\newabf\ by exploiting the
fact that the solutions of the BFKL and GLAP
equations coincide at leading twist if their respective evolution kernels are
related by a ``duality'' relation.  In the fixed coupling limit
the duality relation is simply given by:
\eqn\dualdef{
\chi(\gamma(N,\as),\as)=N.} The splitting function then will contain
 all relative corrections of order
$(\alpha_s \log{1/x})^n$, derived from
$\chi_0(M)$, and of order
$\alpha_s(\alpha_s \log{1/x})^n$, derived from $\chi_1(M)$.

The early wisdom on how to implement the information from $\chi_0$ was
completely shaken by  the computation of
$\chi_1$, which showed that the naive expansion for the
improved anomalous dimension had a hopelessly
bad behaviour. However, as a
consequence of the physical requirement of momentum 
conservation,  
this problem is
cured if the small-$x$ resummation is
combined with the standard resummation of collinear singularities, by
constructing a `double-leading' perturbative
expansion. 

However, higher order corrections to the kernel
qualitatively change the asymptotic small-$x$ behaviour
of structure functions by changing it from
$x^{-\lambda_0}$ to $x^{-\lambda}~=~x^{-\lambda_0} e^{\Delta \lambda
\xi}~\approx ~x^{-\lambda_0}[1+\Delta \lambda \xi+....]$, with:
\eqn\lambdadef{\lambda=\lambda_0+\Delta
\lambda,~~~~~\lambda_0=\as\chi_0(\half)=\as
c_0,~~~~~\Delta\lambda=\as^2\chi_1(\half)+\cdots. }
The
computed correction $\chi_1(\half)$ to the leading result is quite
large, and this suggests that
the
correct asymptotic exponent is not reliably determined by the two
known terms.
Therefore in our previous work we have treated $\lambda$ as
a parameter to be fitted from the data. Good agreement with the HERA
data was found, but at the price of having to sharply fine-tune this parameter.

In our recent work~\newabf\ we have shown that also this problem can
be 
solved by a full treatment of running coupling corrections.
Indeed, it has been known for
some time that
running coupling effects can be included perturbatively order by order at
small $x$ by adding effective subleading
$\Delta \chi_i$ contributions to the BFKL kernels $\chi_1, \chi_2,
\dots$.
However, these additional terms turn out to have singularities at $M=1/2$,
which correspond to an enhancement by powers
of $\ln 1/x$ of the associated splitting
functions, which may offset the perturbative suppression by powers of
$\as$. In our approach, we have shown that these contributions
can be resummed to all orders at the level of splitting functions, in
a way compatible with factorization and  a
 smooth behaviour in the small-$x$ limit.

Based on these results, we now know the way the
information
contained in
$\chi_0(M)$ should be used in order to construct a better first approximation
for the improved anomalous dimension.
Indeed, we find that, once running coupling effects are properly included in
the improved anomalous dimension, the
asymptotic behavior near $x=0$ is much softened with respect to the naive
Lipatov exponent. Hence, the corresponding
dramatic rise of structure functions at small $x$, which is phenomenologically
ruled out, is replaced by a milder rise. The effective
exponent corresponding to this rise
turns out to be in surprisingly good agreement with the fine-tuned
value of $\lambda$ indicated by the fit, which in turn led to a
splitting function which closely followed the NLO GLAP in the region
of the data.
This suggests that additional higher order corrections are small, and 
that a leading--order approximation based on the standard BFKL kernel
$\chi_0$ is phenomenologically viable. 

We discuss now explicitly our proposed improved splitting function.
Assuming that one only knows $\gamma_0(N)$, $\gamma_1(N)$ and
$\chi_0(M)$, 
the improved anomalous dimension has the following expression:
\eqn\leadimprnl{\eqalign{
&\gamma_I^{NL}(\as, N) = [\as\gamma_0(N)+ \as^2 \gamma_1(N) +
\gamma_s(\smallfrac{\as}{N}) -\smallfrac{n_c\as}{\pi N}]+\cr &\qquad
+\gamma_A(c_0,\as,N)-\half +\sqrt{\smallfrac{2}{\kappa_0\as}[N-\as
c_0]}
+\quarter\beta_0\as(1+\frac{\as}{N} c_0)-\rm{mom.~sub.}}} 
The first line on
the right-hand side, within square brackets,
is the 
double-leading expression for the improved anomalous
dimension at this level of accuracy, made up of the NLO perturbative
term
$\as\gamma_0(N)+\as^2\gamma_1(N)$ plus the power series of terms
$(\as/N)^n$ in $\gamma_s(\smallfrac{\as}{N})$, obtained
from $\chi_0$ using eq.~\dualdef, with subtraction of the order
$\as$ term to avoid double counting ($c_A=n_c=3$). 
In the second
line, the ``Airy'' anomalous dimension
$\gamma_A(c_0,\as,N)$ contains the running coupling resummation, 
and the remaining terms
subtract the contributions to $\gamma_A(c_0,\as,N)$
which are already included in
$\gamma_s$,
$\gamma_0$ and $\gamma_1$.   
The Airy anomalous dimension $\gamma_A(c_0,\as,N)$ 
is the exact solution of the 
running coupling BFKL equation corresponding to a quadratic 
appoximation of $\chi_0$ near $M=\half$: $\chi_0 \approx
[c_0+\half\kappa_0(M-\half)^2]$. Finally ``$\rm{mom.~sub.}$" is a
subleading subtraction  that ensures momentum
conservation
$\gamma_{I}(\as,N=1)=0$. 

We now summarize the properties of the improved anomalous dimension in this
approximation. In the limit $\as
\rightarrow 0$ with arbitrary $N$, $\gamma_I(\as,N)$ reduces to
$\as\gamma_0(N)+\as^2\gamma_1(N)+O(\as^3)$. For  $\as
\rightarrow 0$ with $\as/N$ fixed, $\gamma_I(\as,N)$  reduces to
$\gamma_{\rm DL-LO}=\as\gamma_0(N)+\gamma_s(\smallfrac{\as}{N})
-\smallfrac{n_c\as}{\pi N}$, i.e. the leading term of the
double-leading expansion. Thus the Airy term is subleading in both
limits. In spite of this, its role is very significant because of the
singularity structure of the different terms in
eq.~\leadimprnl. In fact,
$\gamma_0(N)$ has a pole at $N=0$, $\gamma_s$ has a branch cut at
$N=\as c_0$, and
$\gamma_A$ has a pole at $N=N_0<\as c_0$, where $N_0$ is the position of the
rightmost zero of the Airy function. The
importance of the Airy term is that the square root term subtracted from
$\gamma_A$ cancels, within the relevant accuracy, the branch cut of $\gamma_s$
at $N=\as c_0$ and replaces the
corresponding asymptotic behaviour at small $x$ with the much softer one from
$\gamma_A$. Note that the quadratic approximation is sufficient to
give the correct asymptotic behaviour up to terms which
are of higher order in comparison to those included in the double-leading
expression in eq.~\leadimprnl.

\topinsert
\vbox{
\epsfxsize=14truecm
\centerline{\epsfbox{guido1.ps}}
\hbox{
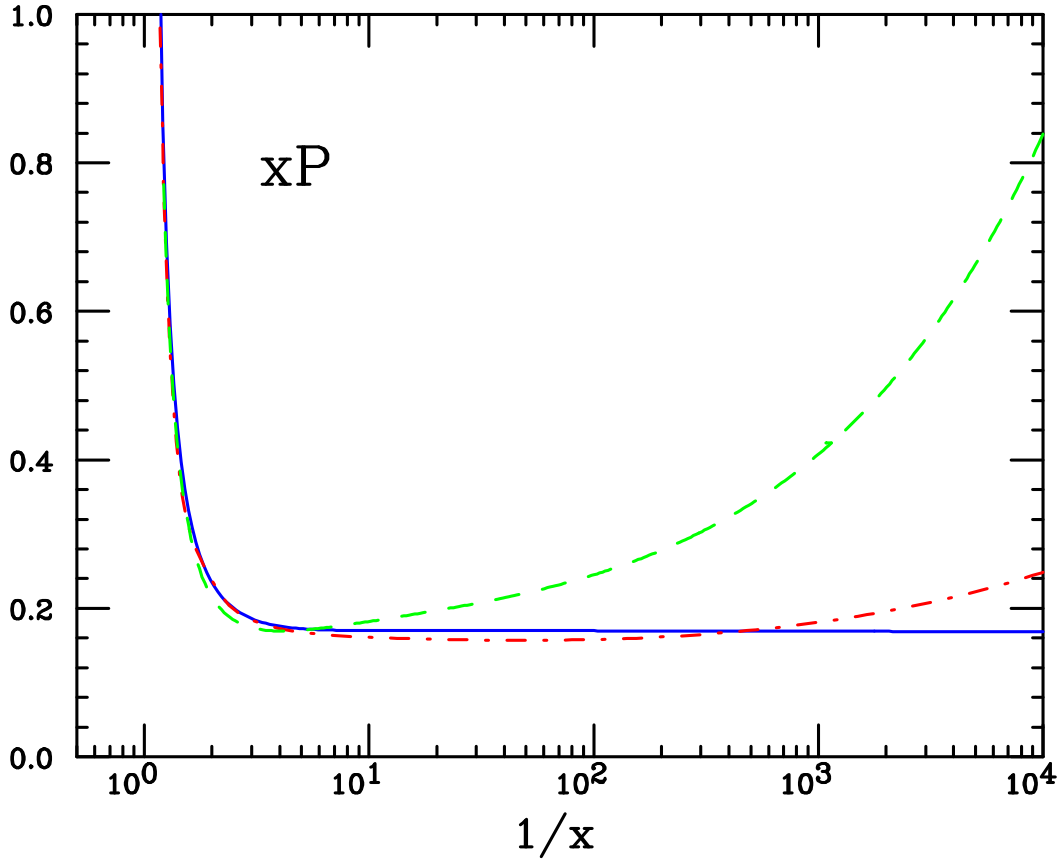
\vbox{\footnotefont\baselineskip6pt\narrower\noindent Figure 1: The improved splitting function corresponding to
$\gamma_I(\as,N)$ eq.~\leadimprnl\ with
$\as=0.2$ (dot-dashed), compared with those from the DL-LO approximation (dashed) and 
GLAP NLO (solid). }}\hskip1truecm}
\endinsert
While we refer the interested reader to ref.~\newabf\ for further details, we 
show here two new figures for the splitting function which illustrate
the success of this approach and compare it both to our older results and
to those of ref.~\newciaf.
In figure 1 we show (for $\as=0.2$) the singlet
splitting function obtained from eq.~\leadimprnl\ 
compared with the NLO GLAP kernel and with the DL-LO approximation,
which displays the sharp small-$x$ rise characteristic of the BFKL resummation.
In the region of the HERA
data,  our improved splitting function,  with no free
parameters, closely follows
the  NLO GLAP evolution with a behaviour at
small $x$ which is much softer than that of BFKL. It is interesting to
note that the agreement between GLAP and resummed results is
significantly improved by the inclusion in eq.~\leadimprnl\ 
of $\gamma_1$ and the
corresponding double-counting subtraction. This improvement was
already shown in ref.~\newabf\ in the anomalous dimension,
and is even more apparent in the splitting function shown here.

\topinsert
\vbox{
\epsfxsize=14truecm
\centerline{\epsfbox{guido2.ps}}
\hbox{
\vbox{\footnotefont\baselineskip6pt\narrower\noindent Figure 2: 
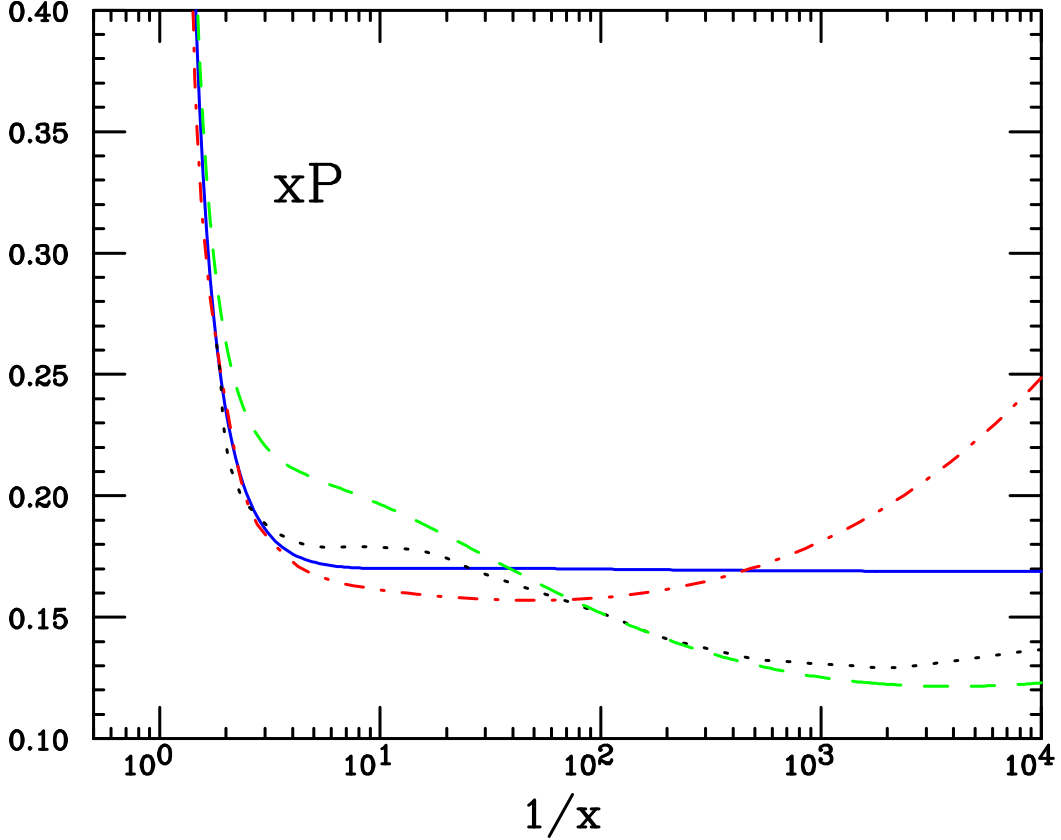
The improved splitting function corresponding to
$\gamma_I(\as,N)$ eq.~\leadimprnl\ with
$\as=0.2$ (dot-dashed), compared with  GLAP NLO (solid), with the
result of our old approach, where the value of
$\lambda=0.21$ was fine tuned to best fit
the HERA data (dashed), and with the splitting function of
ref.~\newciaf\ (dotted). Note that the vertical scale is 
rather larger than that in figure 1. }}\hskip1truecm}
\endinsert

In figure 2  
the improved splitting function  is compared with those  of
our earlier approach to this problem 
and with that recently obtained in ref.~\newciaf. It is seen that
the correct inclusion of   running coupling
corrections is by itself sufficient to produce
the softening of the behaviour at small $x$ required in order to
reproduce the data. 
Thus it is no
longer  necessary to introduce
$\lambda$ by hand as a free parameter. In fact, the fine-tuned value
$\lambda\approx 0.21$  (for
$\as\sim0.2$)
obtained from fitting the data is
 in remarkable agreement with the value $N_0=0.2112$ determined
from the Airy anomalous dimension with $\as = 0.2$. This improved 
splitting function 
is so good that subleading corrections to it are presumably very small.

In the same figure,  we also show the splitting function from
ref.~\newciaf. 
In comparison to the approach adopted there, 
we share the general physical framework, but there are a number of
differences. Specifically, the authors of ref.~\newciaf\
determine the asymptotic small-$x$ behaviour by
making some assumptions on the way to implement a symmetrisation 
$M \rightarrow
(1-M)$ of the BFKL kernel
$\chi(\alpha_s,M)$, which we prefer not to do because of the ambiguities
it introduces. Also, our resummed curve in figure 2 includes the
effects of $\gamma_1$ but not fully those of $\chi_1$. 
This is because we find that after running coupling resummation, the
impact of the left-over terms in $\chi_1$ is comparable to the
resummation ambiguities. 
By contrast, their curve
includes the effects of $\chi_1$ but not those of
$\gamma_1$. Furthermore, in ref.~\newciaf\ 
the running coupling evolution equation is
solved explicitly
but numerically to
the relevant order. Instead, through the Airy expansion, we obtain an
analytic expression which holds to the same accuracy.
This opens up the possibility to  fit the data and match to usual
evolution equations in a straightforward way. The resummed splitting 
functions shown in figure 2  all rise asymptotically with
approximately the same power, but differ in the preasymptotic
region. These differences can be taken as a
measure of the remaining resummation uncertainties. 
\bigskip
\footatend
\immediate\closeout\rfile\writestoppt
\baselineskip=14pt\centerline{{\bf References}}\bigskip{\frenchspacing%
\parindent=20pt\escapechar=` \input refs.tmp\vfill\eject}\nonfrenchspacing

\vfill\eject
\bye

\bigskip
\footatend
\immediate\closeout\rfile\writestoppt
\baselineskip=14pt\centerline{{\bf References}}\bigskip{\frenchspacing%
\parindent=20pt\escapechar=` \input refs.tmp\vfill\eject}\nonfrenchspacing
\vfill\eject
\bye